\documentclass[epj,final]{svjour}
\usepackage{graphics}
\usepackage{amsfonts}
\usepackage{latexsym}

\newcommand{\abs}[1]{\mid \! #1 \! \mid}
\newcommand{\abmtt}{a$_B^{-3}$}
\newcommand{\abmtm}{\textrm{a}_B^{-3}}
\newcommand{\prl}{Phys. Rev. Lett.}
\newcommand{\prb}{Phys. Rev. B}
\newcommand{\pra}{Phys. Rev. A}

\begin{document}

\title{First-Principles Calculation of Electric Field Gradients
and Hyperfine Couplings in YBa$_2$Cu$_3$O$_7$}

\author{S. Renold\inst{1}$^{,\star}$, S. Pliber\v{s}ek\inst{1}, E.P. Stoll\inst{1}, T.A. Claxton\inst{2}, and P.F. Meier\inst{1}.}

\thanks{e-mail: \texttt{sam@physik.unizh.ch}}

\institute{Physics Institute, University of Zurich, CH-8057 Zurich, 
         Switzerland \and
         Department of Chemistry, University of York, York, YO10 5DD, UK}

\date{Received: date / Revised version: date}

\abstract{
The local electronic structure of YBa$_2$Cu$_3$O$_7$ has been calculated using first-principles cluster methods. Several clusters embedded in an appropriate background potential have been investigated. The electric field gradients at the copper and oxygen sites are determined and compared to previous theoretical calculations and experiments. Spin polarized calculations with different spin multiplicities have enabled a detailed study of the spin density distribution to be made and a simultaneous determination of magnetic hyperfine coupling parameters. The contributions from on-site and transferred hyperfine fields have been disentangled with the conclusion that the transferred spin densities essentially are due to nearest neighbour copper ions only with marginal influence of ions further away. This implies that the variant temperature dependencies of the planar copper and oxygen NMR spin-lattice relaxation rates are only compatible with commensurate antiferromagnetic correlations. The theoretical hyperfine parameters are compared with those derived from experimental data.
}

\authorrunning{S. Renold {\it et al.}}

\titlerunning{Electric Field Gradients and Hyperfine Couplings in YBa$_2$Cu$_3$O$_7$}

\maketitle

%---------------------
%---------------------
\section{Introduction}
%---------------------
%---------------------

Although discovered~\cite{bib:bednorzmuller} in 1986, the high temperature superconducting (HTS) copper oxides continue to present challenge to both experiment and theory. The underlying mechanism (or mechanisms) responsible for their superconductivity is still ambiguous. To understand the various contributions of magnetism and phonons in high temperature superconductivity requires a detailed knowledge of the electronic structure of HTS materials. These exhibit considerable structural complexity but this complexity brings the advantage of various nuclear species located throughout the unit cell which serve as local probes of the electronic and magnetic properties. Through nuclear quadrupole and magnetic resonance (NQR and NMR) techniques, these nuclei probe passively the local electronic structure and have delivered a tremendous amount of data about both static and dynamic properties of HTS materials~\cite{bib:reviews}.

The nuclei experience the spin polarization of the electronic system through various hyperfine interactions which basically are well known. It turned out, however, that in HTS copper oxides their analysis is not as straightforward as expected. For example, the spin shifts which give information on the static spin susceptibility have to be separated from the chemical shifts which, being independent on temperature, are assumed to dominate the Knight shifts at low temperature. Precise measurements of these NMR shifts in the superconducting state, however, are difficult due to the field inhomogeneities produced by the vortex lattice. Furthermore, the spin shifts at the planar copper nuclei are extremely small in all cuprates when measured with the applied field perpendicular to the planes.

Dynamic properties of the spin fluid can be probed by the various nuclear relaxation rates. The interaction energy is determined by the sum of the hyperfine interactions of which the transferred hyperfine fields make a crucial contribution as has first been pointed out in Refs.~\mbox{\cite{bib:milarice,bib:shastry}}. In particular, the variant temperature dependence of the spin-lattice relaxation rates at planar copper and oxygen nuclei could be explained by assuming that the transferred fields at $^{17}$O from the two neighbouring copper ions cancel if these moments are antiferromagnetically coupled whereas the copper nuclei could reflect the antiferromagnetic spin fluctuations. Neutron scattering measurements, however, indicated that the fluctuation peaks were incommensurate and not located at the reciprocal wave vectors $Q=(\pi / a, \pi / a)$.

Zha, Barzykin, and Pines~\cite{bib:zha} then advocated transferred hyperfine fields at the oxygens from next nearest neighbour copper ions which would further suppress the contributions from the two nearest neighbours. With this extended model they could explain a variety of NMR data in good agreement with incommensurate antiferromagnetic fluctuations as exhibited by neutron scattering results.

Although the focus of NQR and NMR investigations of HTS materials has recently changed to inhomogeneities, in particular static and dynamic stripe phenomena, some of the ``older'' problems still wait for definitive answers. Among these problems the applicability of a one-band approach to describe the low-lying spin and charge excitations in these materials is of special importance. Several experiments~\mbox{\cite{bib:walstedtshastry,bib:martindale,bib:penningtonyugorny}} indicate that a second spin fluid with independent dynamics is required.

Despite the rich information these experiments provide, there exist only few theoretical first-principles approaches which address the determination of electric field gradients (EFGs) and magnetic hyperfine interactions. For the YBa$_2$Cu$_3$O$_{6+x}$ system, EFGs at various nuclear sites have been early obtained by Das and co-workers~\mbox{\cite{bib:das,bib:das1,bib:das2}} with {ab initio} cluster calculations using the unrestricted Hartree-Fock (UHF) method, and  by Schwarz and co-workers~\mbox{\cite{bib:schwarz1,bib:schwarz2}} and Yu {\it et al.}~\cite{bib:yu}, who employed the full-potential linear augmented-plane-wave method (FLAPW) within the local density approximation (LDA).
Results of Hartree-Fock (HF) calculations have also been published by Winter~\cite{bib:winter}. The EFGs calculated with these three different methods agree, more or less, and reproduce apart from one exception the experimental data quite satisfactorily. The exceptional case is the EFG at the planar Cu sites. Using a large cluster that contained 74 atoms, we performed~\cite{bib:physicac} calculations with the density functional (DF) method and obtained values for the Cu EFGs which were in better agreement with the experiments.
These calculations were spin non-polarized and no information on magnetic hyperfine interactions was available. Spin polarized calculations, however, have recently been performed~\cite{bib:physrevb} for the La$_2$CuO$_4$ system which allowed the study of hyperfine coupling parameters.

In this paper, we report the results of extensive cluster studies of the electronic structure of YBa$_2$Cu$_3$O$_7$. Spin polarized calculations with the DF method with generalized gradient corrections to the correlation functionals have been performed for various clusters. The resulting electronic structure, the charge and spin distributions have been analyzed.
In particular, the on-site and transferred hyperfine fields have been investigated in detail and it is demonstrated, that transferred interactions from copper ions beyond the nearest neighbours are small for both planar copper and oxygen nuclei. The relevance of this finding for the interpretation of $^{17}$O NMR spin-lattice relaxation is discussed.

In Sec.~\ref{sec:clu_bas} the clusters used in this work are described together with the basis sets and the methods of the calculations. We then consider in Sec.~\ref{sec:cuo5_ion} first a (CuO$_5$)$^{8-}$ ion and compare the results with predictions for a Cu$^{2+}$ ion in a single-electron approach. The results for the EFGs are given in Sec.~\ref{sec:EFG} and compared to previous theoretical approaches and to experimental values. The magnetic hyperfine interactions are discussed in Sec.~\ref{sec:mag_hyp} where special attention is given to study the transferred fields. The mechanism of spin transfer is analyzed in detail and the calculated hyperfine couplings are compared with values derived from experiments. Sec.~\ref{sec:conclusion} contains a summary and the conclusions. 

%-----------------------------------------
%-----------------------------------------
\section{Clusters and theoretical methods}
%-----------------------------------------
%-----------------------------------------

\label{sec:clu_bas}

The general idea of the cluster approach to electronic structure calculations of properties which depend predominantly on local structures is that the parameters that characterize a small cluster can be transferred to the solid and largely determine its behaviour. One should expect reliable results for local quantities like the EFG or the magnetic hyperfine fields. It would be much more questionable to calculate quantities which are determined only by a large region of the solid. 

It would be desirable to have a cluster that contains as many atoms as possible, but there are two computational limitations to the cluster size: the available computer resources and the convergence of the minimization procedure. The rest of the solid must be simulated by a shell of neighbouring pseudopotentials and, beyond that, point charges so that the cluster experiences the correct electrostatic potential. It is necessary, however, that the results obtained should be checked with respect to their dependence on the cluster size.

An important feature of the cluster approach is that the atom to consider should be in a position of the cluster which respects as much as possible its natural surrounding. This means that the cluster should retain the ``local symmetry'' around that atom. Since we are interested in the EFG of both the planar copper and the chain copper, we used clusters of two different symmetries according to which copper atom we were interested in. The alternative of choosing a single cluster to look at those two ions is currently prohibitive. However both these clusters include nearest neighbour atoms of the CuO$_2$ planes, the latter being incorporated to reflect their proven importance.

\begin{figure}[htb]
\resizebox{0.48\textwidth}{!}{%
  \includegraphics{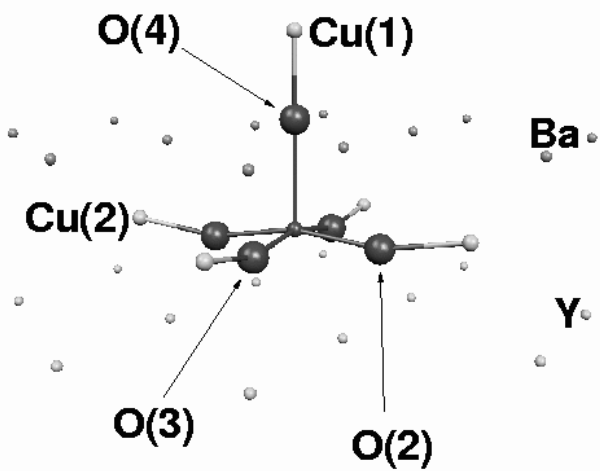}
}
\caption{The CuO$_5$/Cu$_5$Y$_{12}$Ba$_{12}$ cluster centred at a planar copper, Cu(2). Ions with a full basis set are drawn darkly whereas ions with a pseudopotential are drawn lightly.}
\label{fig:cuo5}
\end{figure}

All clusters used assume the low temperature orthorhombic structure of YBa$_2$Cu$_3$O$_7$, with lattice constants $a=3.827$ \AA, $b=3.882$ \AA\ and $c=11.682$ \AA\ as given in Ref.~\cite{bib:unitcell}. We adopt the standard labelling of the copper and oxygen ions as defined in Fig.~\ref{fig:cuo5}. Using the planar Cu(2) atom as a common centre, the crystallographic $a$-axis points towards O(2), the $b$-axis points towards O(3), and the $c$-axis points towards O(4).

The smallest cluster used (see Fig.~\ref{fig:cuo5}) consists of one Cu(2) in the centre surrounded by four oxygen atoms (O(2) and O(3)) of the CuO$_2$ plane and one O(4) (which is called apical oxygen). Surrounding that central part, there are four pseudopotentials in the CuO$_2$ plane with a valence of $+2$ simulating the core electrons of Cu(2) and one pseudopotential for the chain copper, Cu(1), with a valence of $+1$, twelve pseudopotentials for the Ba$^{2+}$ ions above the plane and twelve pseudopotentials for the Y$^{3+}$ beneath the plane. These pseudopotentials screen the electrons of the cluster from the naked positive point charges. The cluster is labelled CuO$_5$/Cu$_5$Y$_{12}$Ba$_{12}$. In addition, these 35 atoms are surrounded by more than 6500 point charges. The positions of some charges at sites far from the cluster centre were adjusted in such a way that the correct Madelung potential in the central region of the cluster was reproduced.

\begin{figure}[htb]
\resizebox{0.48\textwidth}{!}{
\includegraphics{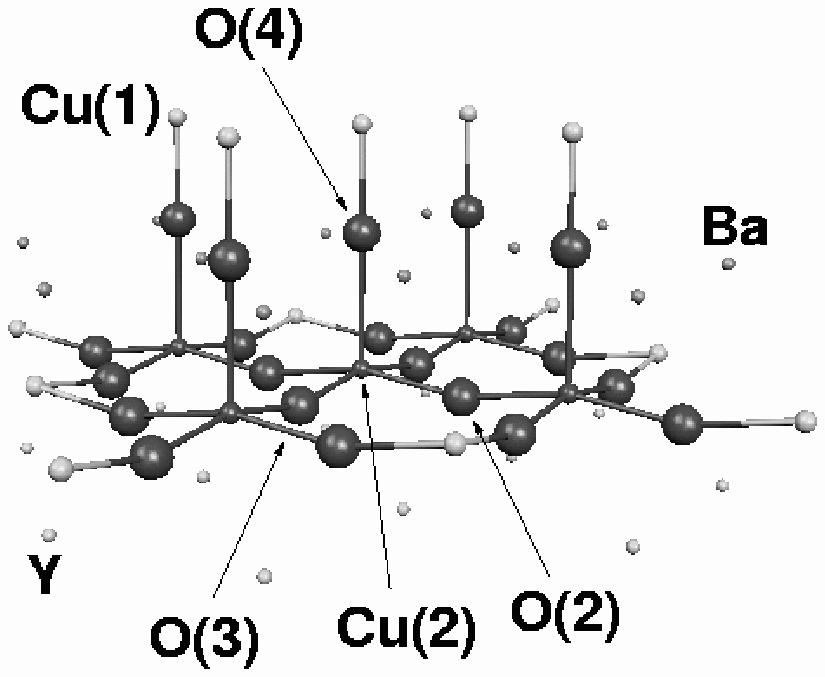}
}
\caption{The Cu$_5$O$_{21}$/Cu$_{13}$Y$_{12}$Ba$_{12}$ cluster centred at a Cu(2).}
\label{fig:cu5o21}
\end{figure}

The largest cluster (see Fig.~\ref{fig:cu5o21}) with a Cu(2) in the centre comprises in total 26 atoms all supporting a basis set; five Cu(2), sixteen planar and five apical oxygens. This cluster is surrounded by a shell of eight Cu(2), five Cu(1), twelve Y and twelve Ba pseudopotentials. That leads to the Cu$_5$O$_{21}$/Cu$_{13}$Y$_{12}$Ba$_{12}$ cluster.

\begin{figure}[htb]
\begin{center}
\resizebox{0.45\textwidth}{!}{
  \includegraphics{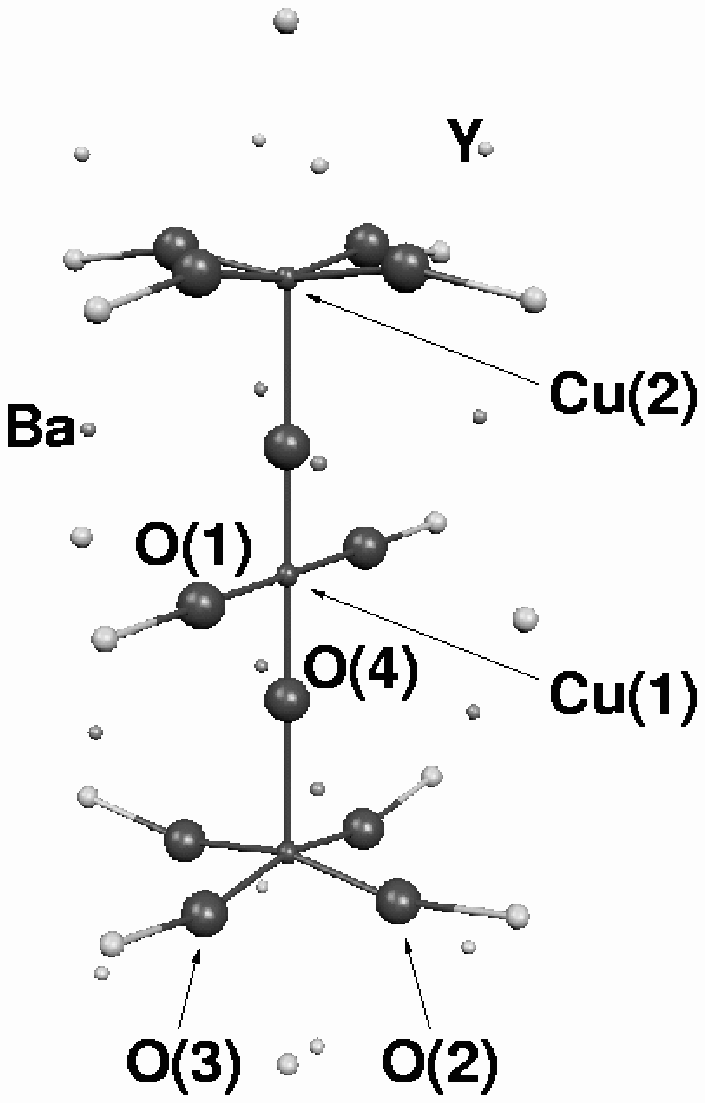}
}
\end{center}
\caption{The Cu$_3$O$_{12}$/Cu$_{14}$Y$_8$Ba$_8$ cluster centred at a Cu(1).}
\label{fig:cu3o12}
\end{figure}

To investigate the chain copper EFG it has been necessary to construct another cluster where a Cu(1) ion is in the centre. In detail the cluster contains 16 atoms, all supporting a basis set; one Cu(1), two Cu(2), two chain O(1), two apical O(4) and eight planar oxygens (four O(2) and four O(3)). There are also four Cu(1), ten Cu(2), eight Ba and eight Y bare pseudopotentials. Fig.~\ref{fig:cu3o12} shows the geometry of this cluster which is labelled Cu$_3$O$_{12}$/Cu$_{14}$Y$_8$Ba$_8$.

Table~\ref{tbl:compilation_I} contains a concise description of each cluster.

In view of the large range of cluster sizes used in this paper we continue our policy~\cite{bib:physrevb} of using the same basis set for each cluster, so that direct comparison can be made, and as before use the 6-311G basis set~\cite{bib:g98}. In any case, as earlier work has shown~\cite{bib:physrevb}, there is no significant change if the quality of this basis set is improved. Pseudopotentials were used~\cite{bib:pseudopotentials} on \mbox{Ba$^{2+}$}, \mbox{Y$^{3+}$}, and those \mbox{Cu$^{2+}$} and \mbox{Cu$^+$} ions indicated in Figs.~\ref{fig:cuo5}, \ref{fig:cu5o21}, and \ref{fig:cu3o12}.
The total number of atoms, electrons, basis functions and primitive Gaussian functions for each cluster is collected in Table~\ref{tbl:compilation_II}.

\begin{table*}[htb]
\caption{Compilation of the used clusters with regard to their constituents.}
\begin{center}
\begin{tabular}{lc|ccccc|cccc}
\hline
&& 
\multicolumn{5}{c|}{atoms with a full basis set} &
\multicolumn{4}{c}{atoms with pseudopotentials}  \\ \hline
cluster & centre & Cu(2) & Cu(1) & O(2,3) & O(4) & O(1) & Cu(2) & Cu(1) & Y & Ba \\
CuO$_5$/Cu$_5$Y$_{12}$Ba$_{12}$             & Cu(2) & 1 &   & 4  & 1 &   & 4 & 1 & 12 & 12 \\
Cu$_5$O$_{21}$/Cu$_{13}$Y$_{12}$Ba$_{12}$   & Cu(2) & 5 &   & 16 & 5 &   & 8 & 5 & 12 & 12 \\
Cu$_3$O$_{12}$/Cu$_{14}$Y$_8$Ba$_8$         & Cu(1) & 2 & 1 & 8  & 2 & 2 & 4 &10 & 8  & 8  \\ \hline
\end{tabular}
\end{center}
\label{tbl:compilation_I}
\end{table*}

In contrast to our previous work~\cite{bib:physicac} on YBa$_2$Cu$_3$O$_7$ we present here \emph{spin polarized} calculations using the Gaussian 98 software package~\cite{bib:g98}.

We showed in an earlier work~\cite{bib:physrevb} that the density functional (DF) method, a model which includes correlation effects, describes the covalent Cu-O bonding better than Hartree-Fock (HF) methods. Although the DF operator, equivalent to the Hamilton operator in the HF method, is not uniquely defined, the different exchange and correlation functionals available give effectively equivalent results. In addition several forms of the gradient correction of these functionals are available. So to maintain a consistent calculational scheme, following on from our work~\cite{bib:physrevb} on \mbox{La$_2$CuO$_4$}, we persevere with the exchange functional proposed by Becke~\mbox{\cite{bib:becke1,bib:becke2}} coupled with the correlation functional of Lee, Yang, and Parr~\cite{bib:LYP} (specified by the BLYP keyword in the Gaussian 98 program). For the final analysis a special program was written~\cite{bib:stoll97} which used the integrals and electronic wave functions from the Gaussian calculations as data.

\begin{table}[htb]
\caption{Compilation of the used clusters: N: number of atoms, E: number of electrons, B: number of basis functions, P: number of primitive Gaussian functions.}
\begin{center}
\begin{tabular}{lcccc}
\hline
cluster                                   &  N &   E &   B &   P \\ \hline
CuO$_5$/Cu$_5$Y$_{12}$Ba$_{12}$           &  6 &  77 & 104 & 196 \\
Cu$_5$O$_{21}$/Cu$_{13}$Y$_{12}$Ba$_{12}$ & 26 & 345 & 468 & 876 \\
Cu$_3$O$_{12}$/Cu$_{14}$Y$_8$Ba$_8$       & 15 & 200 & 273 & 510 \\ \hline
\end{tabular}
\end{center}
\label{tbl:compilation_II}
\end{table}

To define the notation we specify that the EFG components, $V^{ii}$, are given here in atomic units, i.e.\ in \mbox{$e \abmtm$} $=$ \mbox{$- \abs{e} \textrm{a}_B^{-3}$} $=$ \mbox{$- \textrm{Ha} / \textrm{a}_B^2$}.
The quantity $q^{ii}= V^{ii}/ \abs{e}$ then corresponds to $-9.7174 \times 10^{21}$ in units of V/m$^2$.

A general hyperfine interaction tensor, with components $\alpha \beta$, is denoted by $f^{\alpha \beta}$ and has units of \abmtt, which importantly is independent of the underlying nuclear structure.

The hyperfine interaction energy at nucleus $k$ is then given by

\begin{equation}
^k\!F^{\alpha \beta} = \hbar \ ^k\! \gamma \  \hbar \gamma_e f^{\alpha \beta}
\end{equation}
where $^k\! \gamma$ and $\gamma_e$ are the gyromagnetic ratios for nucleus $k$ and the electron respectively.

We follow the NMR literature in replacing our general $f^{\alpha \beta}$ by $a^{\alpha \beta}$, for the Cu on-site term, $b^{\alpha \beta}$ and $c^{\alpha \beta}$ for the transferred hyperfine fields when it refers to Cu and O, respectively.

We can identify at least three contributions to a general $f^{\alpha \beta}$:

\begin{equation}
f_{tot}^{\alpha\beta} = f_{iso} \delta^{\alpha\beta} + f_{dip}^{\alpha\beta} + f_{so}^{\alpha\beta},
\end{equation}
where $f_{iso}$ is the isotropic (Fermi contact) term, $f_{dip}^{\alpha\beta}$ is the traceless dipolar term, and $f_{so}^{\alpha\beta}$ the spin-orbit coupling. The difference between on-site and transferred contributions will be discussed in Sec.~\ref{sec:mag_hyp}.

%-----------------------------
%-----------------------------
\section{Embedded CuO$_5$ ion}
%-----------------------------
%-----------------------------

\label{sec:cuo5_ion}

As a starting point, we consider the (CuO$_5$)$^{8-}$ ion (see Fig.~\ref{fig:cuo5}) embedded in an appropriate lattice of pseudopotentials and point charges in the YBa$_2$Cu$_3$O$_7$ crystal.
In Table~\ref{tbl:efg_comparison} we collected results for the Mulliken charge $\rho_M$, the components of the EFG $V^{ii}$, the Mulliken spin density $\rho$, the isotropic term $a_{iso}$, and the diagonal elements of the dipolar hyperfine tensor $a_{dip}^{ii}$ at the copper site. A comparison with the values in the La$_2$CuO$_4$ system~\cite{bib:physrevb} shows that the lacking apex oxygen in the YBa$_2$Cu$_3$O$_7$ compound has not a great influence on these values.

\begin{table*}[htb]
\caption{Compilation of results for the Cu(2) in the smallest clusters: a (CuO$_5$)$^{8-}$ ion in the YBa$_2$Cu$_3$O$_7$ system and a (CuO$_6$)$^{10-}$ ion in the La$_2$CuO$_4$ system. The Mulliken charge and the Mulliken spin density attributed to the copper atom are denoted by $\rho_M$ and $\rho$ respectively. The components of the EFG, $V^{ii}$, are given in atomic units. The isotropic term, $a_{iso}$, and the dipolar hyperfine couplings, $a_{dip}^{ii}$, are in units of \abmtt.}
\begin{center}
\begin{tabular}{cccccccccc}
\hline
                   & $\rho_M$ & $V^{xx}$ & $V^{yy}$ & $V^{zz}$ & $\rho$ & $a_{iso}$ & $a_{dip}^{xx}$ & $a_{dip}^{yy}$ & $a_{dip}^{zz}$ \\ \hline
YBa$_2$Cu$_3$O$_7$ & 1.111 & $-$0.582 & $-$0.485 & 1.067 & 0.633 & $-$1.969 & 1.651 & 1.676 & $-$3.327 \\
La$_2$CuO$_4$      & 1.167 & $-$0.706 & $-$0.706 & 1.412 & 0.667 & $-$1.782 & 1.764 & 1.765 & $-$3.529 \\ \hline
\end{tabular}
\label{tbl:efg_comparison}
\end{center}
\end{table*}

We can compare these results for the hyperfine coupling constants to the predictions of a single electron picture of a Cu$^{2+}$ ion. Bleaney {\it et al.}~\cite{bib:bleaney} assign to a single 3d$_{x^2-y^2}$ hole a dipolar coupling constant of

\begin{equation}
a_{dip}^{zz} = -\frac{4}{7} \left< r^{-3} \right>
\label{eq:adipzz}
\end{equation}
and a core polarization of
\begin{equation}
a_{cp} = -\kappa \left< r^{-3} \right>.
\label{eq:acp}
\end{equation}
with a parameter $\kappa=0.26 \pm 0.06$.

In this paper we have preferred to use $a_{iso}$ instead of $a_{cp}$ since we are calculating the total Fermi contact interaction. This could be dominated by $a_{cp}$ but $a_{cp}$ is not an observable. Eq.~(\ref{eq:acp}) however is the correct usage for the model system proposed.

If our dipolar hyperfine coupling $a_{dip}^{zz}=-3.327$~\abmtt\ is inserted into Eq.~(\ref{eq:adipzz}) the estimated value for $\left<r^{-3}\right>$ is 5.82~\abmtt\ which is close to the expected value. Assuming that the isotropic hyperfine coupling is dominated by core polarization we get $a_{iso}=a_{cp}=-1.51$~\abmtt\ from Eq.~(\ref{eq:acp}) which is reasonable when compared to our properly calculated value in Table~\ref{tbl:efg_comparison} of $a_{iso}=-1.97$~\abmtt.
We have used Eqs.~(\ref{eq:adipzz}) and~(\ref{eq:acp}) to get these estimates and therefore imposed the single electron picture of the Cu$^{2+}$ ion on our cluster results. We can question what our estimate for $\left<r^{-3}\right>$ of 5.82~\abmtt\ really means since we are in a covalently bound molecular situation.  In the atomic picture the larger $\left<r^{-3}\right>$ is, the more compact is the associated atomic orbital (AO) and {\it vice versa}.  In a molecular situation we might expect a smaller $\left<r^{-3}\right>$ due to delocalisation of the orbital when involved in a molecular orbital (MO) but this is too simplistic for the following reason.

The operator $r^{-3}$ will sample only the density close to the copper and so should be similar to the atomic value if the electron was not shared with other atoms. This means that the $\left<r^{-3}\right>$ value will be proportional to the probability of finding the electron in that orbital. In theoretical calculations this is simply the square of the MO coefficient of the copper AO. In this context Eq.~(\ref{eq:adipzz}) is probably best rewritten as

\begin{equation}
a_{dip}^{zz}=-\frac{4}{7}n_{atom}\left<r^{-3}\right>_{atom}
\label{eq:mod}
\end{equation}
where $n_{atom}$ is the orbital occupancy. This equation is very useful if, and only if, $\left< r^{-3} \right>_{atom}$ retains its value in a molecular situation. Otherwise Eq.~(\ref{eq:mod})  will have two unknowns but only one item of experimental data.

We have extracted $\left<r^{-3}\right>_{atom}$ from our calculations and find that the value ($\approx$ 8~\abmtt) is significantly larger than the corresponding value given for the Cu$^{2+}$ ion. This is not entirely unexpected because it has long been known that AOs are more contracted in a molecular situation if we use AOs as basis orbitals in MO theory. It was particularly noticeable for hydrogen atoms and also for d-orbitals. This effect will be amplified if bonding reduces the electron density in the AO, that is, makes the atom more positive. Although this has a passing interest it certainly raises a cautionary note here when using Eq.~(\ref{eq:adipzz}) or the modified form~(\ref{eq:mod}). Since no useful conclusions result from our $\left<r^{-3}\right>=5.8$~\abmtt\ derived from Eq.~(\ref{eq:adipzz}) using our theoretical value for $a_{dip}^{zz}$, it follows that no useful conclusion can be drawn from using any experimental value of $a_{dip}^{zz}$ in the same way.

%---------------------------------
%---------------------------------
\section{Electric field gradients}
%---------------------------------
%---------------------------------

\label{sec:EFG}

\begin{figure}[htb]
\resizebox{0.48\textwidth}{!}{%
  \includegraphics{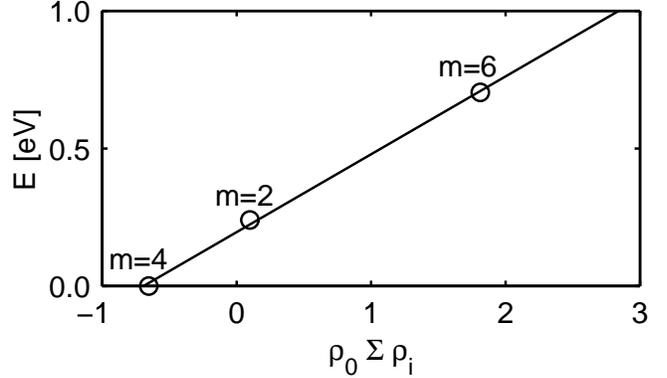}
}
\caption{Dependence of the calculated energy differences $E_{m=6} - E_{m=4}$ and $E_{m=2}-E_{m=4}$ on the Mulliken spin densities.}
\label{fig:exchange}
\end{figure}

To investigate the electronic structure of the CuO$_2$ plane we used the cluster Cu$_5$O$_{21}$/Cu$_{13}$Y$_{12}$Ba$_{12}$ which comprises five Cu$^{2+}$ ions. Calculations were performed for all three spin multiplicities $m=2$, 4, and 6. The lowest energy is obtained for $m=4$ probably reflecting that the spin distribution is reminiscent of an antiferromagnetic system. The corresponding energy differences $E_{m=6} - E_{m=4}$ and $E_{m=2}-E_{m=4}$ are plotted in Fig.~\ref{fig:exchange} versus the product $\rho_0(\rho_1+\rho_2+\rho_3+\rho_4)$ where $\rho_0$ denotes the Mulliken spin density of the central copper ion and $\rho_i$ ($i=1,\ldots,4$) that at the four other copper ions. It is seen that the calculated energies are related to the product of the Mulliken spin densities according to

\begin{equation}
E_m - E_{m=4} = j\rho_0(\rho_1+\rho_2+\rho_3+\rho_4)
\end{equation}
with $j=283$\ meV. This empirical relationship and its connection with the antiferromagnetic exchange interaction will be further discussed elsewhere together with a comparison of the analogous results for other materials.

In Table~\ref{tbl:efgcu2} the calculated values of the EFG components at the central copper atom are reported. The experimental values for the quadrupole resonance frequencies~\cite{bib:pennington} have been converted into EFGs using the value of $Q(^{63}\textrm{Cu})=-0.211$~b for the nuclear quadrupole moment~\cite{bib:sternheimer}. Also given in Table~\ref{tbl:efgcu2} are previous theoretical values obtained from band structure calculations~\mbox{\cite{bib:schwarz1,bib:yu}} and from a spin non-polarized cluster calculation~\cite{bib:physicac}. The present results are in good agreement with the experimental data. Our values slightly depend on the spin multiplicity which reflects the influence of the finite size of the cluster since for $m=4$ the absolute value of the (negative) Mulliken spin density at the central copper ion is smaller than that at the four other copper atoms. This will be further discussed in the next section. A comparison of the EFG values obtained in the small cluster CuO$_5$/Cu$_5$Y$_{12}$Ba$_{12}$ (see Table~\ref{tbl:efgcu2}) and the values at the four non-central copper ions in the large cluster ($V^{zz}=1.41$) shows good convergence with respect to cluster size.
It should be emphasized, however, that the experimentally determined value for the asymmetry parameter ($\eta =0.01$) cannot be accounted for by theoretical calculations which are based on atomic sites determined by crystallographic data. Although the orthorhombicity is small ($(b-a)/b=0.014$) an asymmetry is expected which is appreciably larger than that observed experimentally.

It is noted here that the EFG at the copper nucleus results from relatively large and almost cancelling individual contributions, as is illustrated in Table~\ref{tbl:detailedefg_cupm4} for the example of Cu with spin multiplicity $m=4$. This demonstrates that the theoretical determination of EFGs is quite delicate, as a further detailed analysis~\cite{bib:epstoll_efg} will show. Therefore, it is necessary to describe the total electron density of the atom as accurately as possible, which require the basis sets to be sufficiently large. (In our experience, the 6-311G basis sets~\cite{bib:g98} have proved adequate as has been shown for La$_2$CuO$_4$ where the dependence of the EFG at Cu on the basis set was discussed in detail~\cite{bib:physrevb}.)

A comparison of the theoretical results with experimental values shows that the new spin-polarized calculations here are distinctly better than our previous spin-unpolarized calculations~\cite{bib:physicac} and the FLAPW calculations~\mbox{\cite{bib:schwarz1,bib:yu}}. The FLAPW values are only about 50\% of the experimental values although the contributions from the 3p and 3d orbitals (with the exception of the 3d$_{x^2-y^2}$ orbital) are almost identical with our current calculations. The exception is crucial since according to our calculations the 3d$_{x^2-y^2}$ orbital
is significantly involved in covalent bonding with the 2p$_{\sigma}$ orbitals on the nearest neighbour planar oxygen atoms. The bonding to the apical oxygens is much more ionic.
Clusters are basically molecular ions surrounded by an environment of atomic ions represented by pseudopotential functions or, more remotely, point charges. The wave functions are determined by a molecular orbital approach which is ideally suited to the study of systems which are significantly covalently bonded. Augmented plane wave methods are more suited to systems which are mainly held together by ionic or metallic bonds. For this reason we believe that the cluster calculations are more appropriate than the FLAPW method for the study of local properties in these crystals.

The values of the EFG components at the planar oxygen sites O(2) and O(3) are collected in Table~\ref{tbl:efgo23} together with previous theoretical values and experimental data for resonance frequencies which have been converted into EFGs using the value $Q(^{17}\textrm{O})=-0.02556$~b. The agreement between theory and experiment is good.

The cluster Cu$_3$O$_{12}$/Cu$_{14}$Y$_8$Ba$_8$ was used to investigate the electronic structure of the apical oxygen O(4) and the chain copper Cu(1) with spin multiplicities $m=3$ and $m=5$. The experimentally determined EFG components at Cu(1) exhibit the peculiar feature that the asymmetry parameter $\eta$ is almost equal to 1 which is not due to a particular symmetry of the lattice site.
Our cluster calculations with $m=5$ fail badly to reproduce this asymmetry but those with $m=3$ give $\eta = 0.995$ as shown in Table~\ref{efgcu1}. For the latter, the spin densities reside mainly in the two planes with small values on the Cu(1) and the O(4). Nevertheless, the interplanar exchange coupling is of general interest. A careful study of this problem would, however, require much larger clusters (at least 11 copper ions) which are part of our plans for future investigations.

For the apical oxygen, O(4), we obtain $V^{zz}=-1.61$ and $\eta=0.22$ also in reasonable agreement with experiments~\cite{bib:takigawa} ($\abs{V^{zz}}=1.22$ and $\eta=0.32$) in consideration of the small, but carefully chosen, cluster.

\begin{table*}[h,t,b] \centering
\caption{Comparison of the EFG components in atomic units at the central Cu(2) in the cluster Cu$_5$O$_{21}$/Cu$_{13}$Y$_{12}$Ba$_{12}$ with other theoretical approaches and with experiment.}
\begin{tabular}{lccccc}
\hline
           & multiplicity & $V^{xx}$ & $V^{yy}$ & $V^{zz}$ & $\eta$ \\ \hline
this work  & 2   & $-$0.778   & $-$0.382   &  1.160   & 0.341 \\
           & 4   & $-$0.596   & $-$0.642   &  1.238   & 0.037 \\
           & 6   & $-$0.689   & $-$0.528   &  1.217   & 0.132 \\ \hline
other theoretical results
$^a$   &     & $-$0.841   & $-$0.823 & 1.664 & 0.01 \\
$^b$   &     & $-$0.31    & $-$0.27  &  0.58 & 0.1  \\
$^c$   &     & $-$0.30    & $-$0.29  &  0.59 & 0.02 \\ \hline
Experiment
$^d$   & & $-$0.643 $\pm$ 0.002  & $-$0.628 $\pm$ 0.002 &  1.271 $\pm$ 0.002 & 0.01 \\ \hline
\multicolumn{6}{l}{$^a$Ref.~\cite{bib:physicac}, $^b$Ref.~\cite{bib:schwarz1}, $^c$Ref.~\cite{bib:yu}, $^d$Ref.~\cite{bib:pennington}.} \\ \hline
\end{tabular}
\label{tbl:efgcu2}
\end{table*}

\begin{table}[htb]
\caption{Contributions of on-site AOs with spin up ($\alpha$) and spin down ($\beta$) to the EFG in the $z$-direction for the central copper and their sum calculated with spin multiplicity $m=4$. The remainder gives the contribution from nuclear point charges and AOs centred at other atomic sites.}
\begin{center}
\begin{tabular}{lrrr}
\hline
   & $V^{zz}_{\alpha}$    &  $V^{zz}_{\beta}$    &   $V^{zz}$   \\
 \hline
 remainder      & $ 0.059$ & $ 0.107$  &    0.168  \\
 $p_x$          & $-0.473$ & $-0.437$  &  $-0.910$ \\
 $p_y$          & $-0.506$ & $-0.469$  &  $-0.975$ \\
 $p_z$          & $ 0.357$ & $ 0.337$  &    0.694  \\
 $d_{3z^2-r^2}$ & $ 4.330$ & $ 4.362$  &  $ 8.692$ \\
 $d_{zx}$       &   2.252  & $ 2.225$  &  $ 4.477$ \\
 $d_{yz}$       & $ 2.254$ & $ 2.220$  &  $ 4.474$ \\
 $d_{x^2-y^2}$  & $-2.218$ & $-4.042$  &  $-6.260$ \\
 $d_{xy}$       & $-4.554$ & $-4.566$  &  $-9.120$ \\ \hline
 Total          & $ 1.501$ & $-0.262$  &    1.239  \\ \hline
\end{tabular}
\end{center}
\label{tbl:detailedefg_cupm4}
\end{table}

\begin{table*}[htb]
\caption{Comparison of the EFG components in atomic units at the planar O(2) and O(3) in the cluster Cu$_5$O$_{21}$/Cu$_{13}$Y$_{12}$Ba$_{12}$ with other theoretical approaches and with experiment.}
\begin{center}
\begin{tabular}{lc|cccc|cccc}
\hline
&              & \multicolumn{4}{c|}{O(2)} & \multicolumn{4}{c}{O(3)} \\ \hline
& multiplicity & $V^{xx}$ & $V^{yy}$ & $V^{zz}$ & $\eta$ & $V^{xx}$ & $V^{yy}$ & $V^{zz}$ & $\eta$ \\
this work & 2         & $-$1.347 & 0.772  & 0.575 & 0.15 &  0.718 & $-$1.229  & 0.511 & 0.17 \\
          & 4         & $-$1.341 & 0.764  & 0.577 & 0.14 &  0.763 & $-$1.351  & 0.588 & 0.13 \\
          & 6         & $-$1.101 & 0.684  & 0.417 & 0.24 &  0.662 & $-$1.098  & 0.436 & 0.21 \\ \hline
other$^a$      && $-$1.21  & 0.72   & 0.49  & 0.2  &  0.72  & $-$1.22   & 0.50  & 0.2 \\
$^b$           && $-$1.393 & 0.849  & 0.544 & 0.22 &  0.852 & $-$1.415  & 0.564 & 0.2 \\ \hline
experiment$^c$ & & $\mp$ 1.08 & $\pm$ 0.65 & $\pm$ 0.42 & 0.2 & $\pm$ 0.65 & $\mp$ 1.05 & $\pm$ 0.40 & 0.2 \\ \hline
\multicolumn{10}{l}{$^a$Ref.~\cite{bib:schwarz1}, $^b$Ref.~\cite{bib:yu}, $^c$Ref.~\cite{bib:takigawa}.} \\ \hline
\end{tabular}
\end{center}
\label{tbl:efgo23}
\end{table*}

\begin{table*}[htb]
\caption{Comparison of the EFG components in atomic units at the chain copper in the cluster Cu$_3$O$_{12}$/Cu$_{14}$Y$_{12}$Ba$_{12}$ with other theoretical approaches and with experiment.}
\begin{center}
\begin{tabular}{lccccc}
\hline
& multiplicity & $V^{xx}$ & $V^{yy}$ & $V^{zz}$ & $\eta$ \\
this work & 3 &           0.601 & $-$0.603 & 0.001 & 0.995 \\ \hline
other
$^a$ & & 0.69  & $-$0.76  & 0.07  & 0.8 \\
$^b$ & & 0.565 & $-$0.631 & 0.066 & 0.79 \\
$^c$ & & 0.754 & $-$0.733 & 0.022 & 0.94 \\
$^d$ & & 0.404 & $-$1.153 & 0.749 & 0.3  \\ \hline
experiment
$^e$ & &  0.767 $\pm$ 0.003 &  $-$0.773 $\pm$ 0.003 &  0.006 $\pm$ 0.001 & 0.99 \\ \hline
\multicolumn{6}{l}{$^a$Ref.~\cite{bib:schwarz1}, $^b$Ref.~\cite{bib:yu}, $^c$Ref.~\cite{bib:physicac}, $^d$Ref.~\cite{bib:das1}, $^e$Ref.~\cite{bib:pennington}.} \\ \hline
\end{tabular}
\end{center}
\label{efgcu1}
\end{table*}

%-------------------------------------
%-------------------------------------
\section{Magnetic hyperfine couplings}
%-------------------------------------
%-------------------------------------

\label{sec:mag_hyp}

\subsection{Isotropic contributions}

In the previous section, we have shown that our wave functions can reasonably well predict the EFGs at various sites. Therefore we feel sufficiently confident to calculate other quantities from these wave functions, focusing in this section particularly on the magnetic hyperfine interactions in the CuO$_2$ plane.

The magnetic hyperfine interaction tensor at a certain nucleus can be decomposed into an isotropic part $D$ and a traceless dipolar part $T$, the latter being discussed in the next subsection. The former is given by the difference between the spin-up and spin-down density at the nuclear site $R$

\begin{equation}
D(\vec{R}) = \frac{8 \pi}{3} \left( \sum_m \abs{\psi_m^{\uparrow}(\vec{R})}^2
- \sum_{m'} \abs{\psi_{m'}^{\downarrow}(\vec{R})}^2 \right)
\end{equation}
where the sums extend over all occupied MOs.

In a restricted open-shell calculation, the sum \mbox{$\sum_{m'}$} \mbox{$\abs{\psi_{m'}^{\downarrow}(\vec{R})}^2$} would exactly cancel the partial sum \mbox{$\sum_{m'}$} \mbox{$\abs{\psi_{m'}^{\uparrow}(\vec{R})}^2$} counting over the number of spin-down MOs only and $D(R)$ would be determined solely by the remaining (open-shell) MOs with spin up. In this case, $D(R)$ corresponds to the Fermi contact term which is produced by AOs that are spherically symmetric around $R$. For unrestricted open-shell calculations, as they are discussed here, all MOs with spin up differ from those with down spin. If there are no singly occupied MOs with dominantly s-like AO contribution, as for example in the (CuO$_5$)$^{8-}$ ion discussed above, the quantity $D(R)$ is interpreted as core polarization. In the present context, we have both kinds of contributions. In a first approximation, the MOs with mostly 3d$_{x^2-y^2}$-character are singly occupied and yield a core polarization contribution to $D(\mathrm{Cu})$. Neighbouring copper ions produce in addition transferred hyperfine fields which produce a Fermi contact interaction from the 4s-orbitals and a much smaller contribution from the inner 3s-, 2s-, and 1s-orbitals by spin polarization, in other words, a core polarization. We therefore prefer to call $D(R)$ just the isotropic hyperfine density at the nuclear site $R$.

\begin{figure}[htb]
\resizebox{0.48\textwidth}{!}{%
  \includegraphics{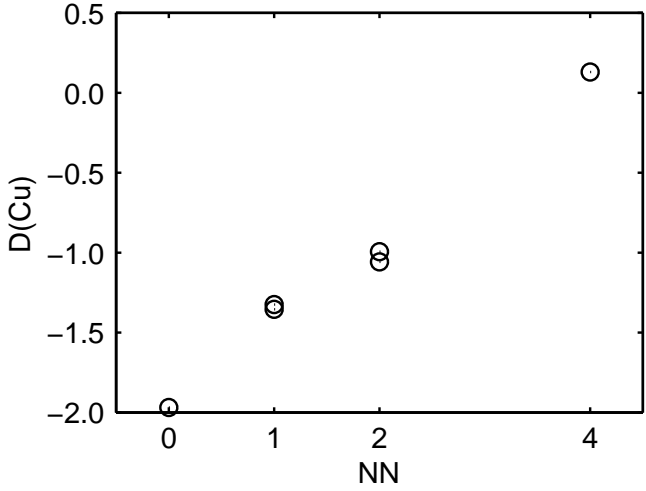}
}
\caption{Isotropic hyperfine density (in units of~\abmtt) at the central copper nucleus plotted against number of nearest copper neighbours.}
\label{fig:dcu}
\end{figure}

The cluster calculations allow us to scrutinize the on-site and transferred hyperfine field in detail. We first focus on the analysis of $D(\mathrm{Cu})$ using results from calculations with the highest possible spin multiplicity.
Thus in the cluster Cu$_5$O$_{21}$/Cu$_{13}$Y$_{12}$Ba$_{12}$ with $m=6$ we have $D(\mathrm{Cu})=0.13$~\abmtt\ at the central copper nucleus, but $D(\mathrm{Cu})=-1.33$~\abmtt\ at the four other copper nuclei. For the small cluster CuO$_5$/Cu$_5$Y$_{12}$Ba$_{12}$, we got $D(\mathrm{Cu})=-1.97$~\abmtt. Furthermore, we have replaced copper ions in the cluster Cu$_5$O$_{21}$/Cu$_{13}$Y$_{12}$Ba$_{12}$ by pseudopotentials without basis sets. This allows us to determine $D(\mathrm{Cu})$ when there are two nearest neighbour (NN) copper ions. The values for $D(\mathrm{Cu})$ obtained in this way are plotted in Fig.~\ref{fig:dcu} as a function of the number of NN.

It is seen, that the isotropic hyperfine density is proportional to the number of NN. In the same way it is found that $D(\mathrm{O})$ at the planar oxygen site is about 0.6~\abmtt\ when there is one NN and about 1.2~\abmtt\ for oxygens between two copper ions.
The sensitivity of the calculated spin densities can be assessed by an inspection of the two values obtained for $D$(Cu) with $\textrm{NN}=1$ or $\textrm{NN}=2$ in Fig.~\ref{fig:dcu}. The slightly different values are due to the small orthorhombicity: the hyperfine density at the central copper nucleus in a cluster with three coppers aligned along the crystallographic direction $a$ amounts to $-0.99$~\abmtt\  but to $-1.06$~\abmtt\ for direction $b$. 

\begin{figure}[htb]
\resizebox{0.48\textwidth}{!}{%
  \includegraphics{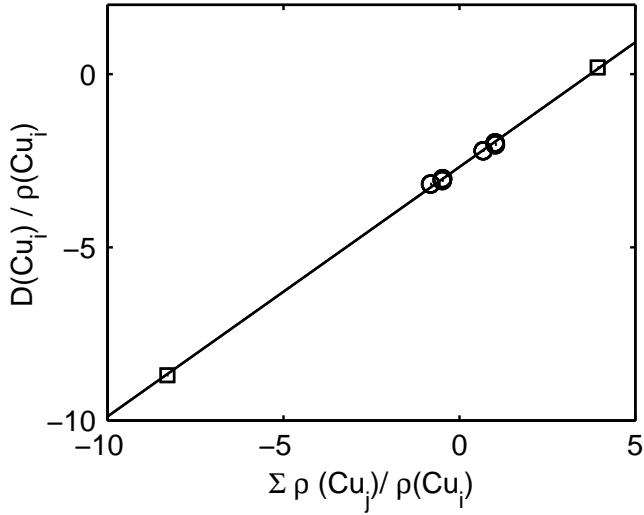}
}
\caption{$D(\textrm{Cu}_i) / \rho(\textrm{Cu}_i)$ plotted against $\sum_{j \in NN}$ $\rho(\textrm{Cu}_j)$ / $\rho(\textrm{Cu}_i)$. The different markers correspond to symmetrically inequivalent copper sites in the cluster which differ by their number of nearest neighbours.}
\label{fig:dcu/rho}
\end{figure}

In the same way we can analyze the results of calculations with lower spin multiplicities. A careful analysis of all results shows that the isotropic hyperfine coupling can well be explained by the following ansatz:

\begin{equation}
{ D(\textrm{Cu}_i) = \alpha_{iso} \rho(\textrm{Cu}_i) + \beta_{iso} \sum_{j \in NN} \rho(\textrm{Cu}_j). }
\end{equation}
Plotting $D(\textrm{Cu}_i) / \rho(\textrm{Cu}_i)$ against $\sum_{j \in NN} \rho(\textrm{Cu}_j) / \rho(\textrm{Cu}_i)$ should then result in a straight line with slope $\beta_{iso}$ and intersection $\alpha_{iso}$. This is done in Fig.~\ref{fig:dcu/rho}.

Similarly the isotropic hyperfine density at an oxygen site can be written as

\begin{equation}
D(\textrm{O}) = \gamma_{iso} \sum_{j \in NN} \rho(\textrm{Cu}_j).
\end{equation}

\begin{figure}[htb]
\resizebox{0.48\textwidth}{!}{%
  \includegraphics{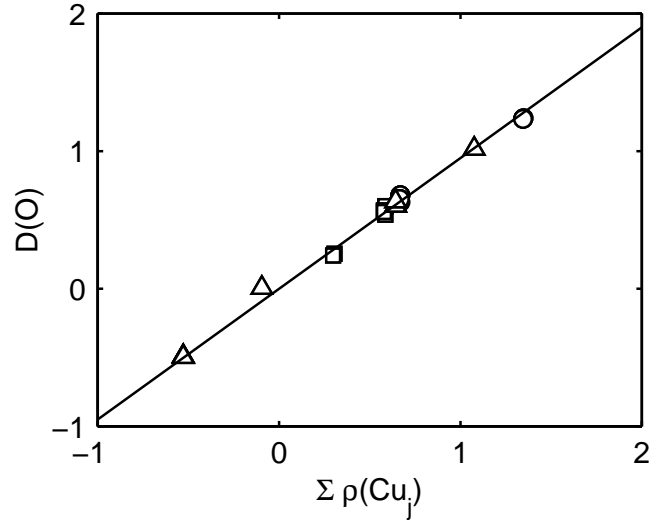}
}
\caption{Isotropic hyperfine density at the planar oxygen nuclei (in units of \abmtt) plotted against $\sum_{j \in NN} \rho(\textrm{Cu}_j)$. The three different markers indicate different spin multiplicities: $\circ$ denotes $m=6$, $\Box$ denotes $m=4$, and $\triangle$ denotes $m=2$.}
\label{fig:do}
\end{figure}

Figure~\ref{fig:do} displays the obtained values in a plot similar to the copper case. Again, the quality of the ansatz can be estimated by noting that the data points fit quite well to a straight line with slope $\gamma_{iso}$. From these two figures it seems quite clear that the influence of further distant atoms can be considered small. To make things more quantitative we have assumed the isotropic hyperfine density at copper and oxygen to depend as well on next nearest neighbours (NNN) and further distant copper ions according to

\begin{displaymath}
\nonumber
D(\textrm{Cu}_i) = \alpha_{iso} \rho(\textrm{Cu}_i) + \beta_{iso} \sum_{j \in NN} \rho(\textrm{Cu}_j) +
\nonumber
\end{displaymath}
\begin{equation}
      +  \beta'_{iso} \sum_{j \in NNN} \rho(\textrm{Cu}_j)
\end{equation}
at a copper nucleus and

\begin{displaymath}
D(\textrm{O}) = \gamma_{iso} \sum_{j \in NN} \rho(\textrm{Cu}_j) + \gamma'_{iso} \sum_{j \in NNN} \rho(\textrm{Cu}_j) + \nonumber
\end{displaymath}
\begin{equation}
                + \gamma''_{iso} \sum_{j \in NNNN} \rho(\textrm{Cu}_j)
\end{equation}
at an oxygen nucleus. A least squares fit to all data yields $\alpha_{iso}=-2.68 \pm 0.02$, $\beta_{iso}=0.72 \pm 0.01$, and $\beta'_{iso} = -0.03 \pm 0.02$ for the constants concerning the copper nucleus and $\gamma_{iso}=0.95 \pm 0.01$, $\abs{\gamma'_{iso}}< 0.025$, and $\abs{\gamma''_{iso}} < 0.03$ for those concerning the oxygen nucleus (in units of \abmtt).

\begin{figure}[htb]
\resizebox{0.48\textwidth}{!}{%
  \includegraphics{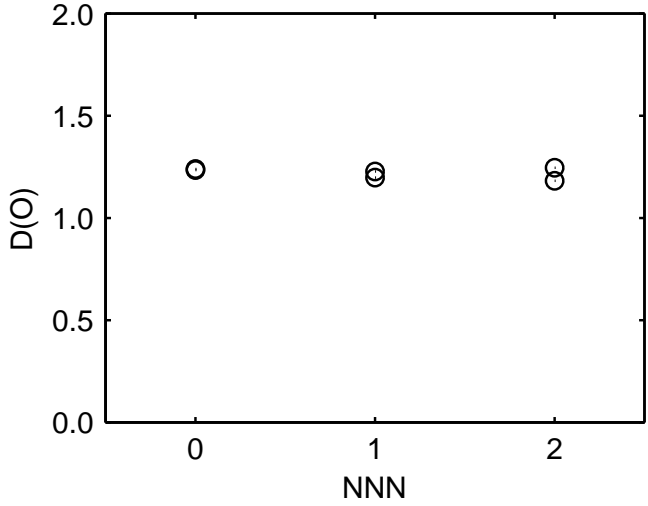}
}
\caption{Isotropic hyperfine density (in units of~\abmtt) at the planar oxygen nuclei as a function of the number of next nearest neighbour copper ions.}
\label{fig:d_nnn}
\end{figure}

It should be noted, that the marginal contribution of NNN ions to $D(\mathrm{O})$ is of particular relevance for the interpretation of the $^{17}$O NMR relaxation rates. To demonstrate this small contribution we plot in Fig.~\ref{fig:d_nnn} the values of $D(\mathrm{O})$ as a function of NNN copper ions. The implications of this result will be discussed further in Sec.~\ref{sec:conclusion}.

%--------------------------------------
%--------------------------------------
\subsection{Dipolar hyperfine coupling}
%--------------------------------------
%--------------------------------------

The dipolar hyperfine coupling results from a spatial average of $1/r^3$ with wave functions:

\begin{displaymath}
T^{\alpha \beta}(\vec{R}) = \sum_{m}   \left< \psi_m^{\uparrow}(\vec{r}) \mid \Delta^{\alpha \beta}(\vec{r}-\vec{R})
\mid \psi_m^{\uparrow}(\vec{r}) \right> - 
\end{displaymath}
\begin{equation}
- \sum_{m'} \left< 
\psi_{m'}^{\downarrow}(\vec{r}) \mid \Delta^{\alpha \beta}(\vec{r}-\vec{R}) \mid \psi_{m'}^{\downarrow}(\vec{r}) \right> 
\end{equation}
with

\begin{equation}
\Delta^{\alpha \beta}(\vec{r})=\frac{3x^\alpha x^\beta - \delta^{\alpha \beta} \abs{\vec{r}}^2}{\abs{\vec{r}}^5}.
\end{equation}

\begin{figure}[htb]
\resizebox{0.48\textwidth}{!}{%
  \includegraphics{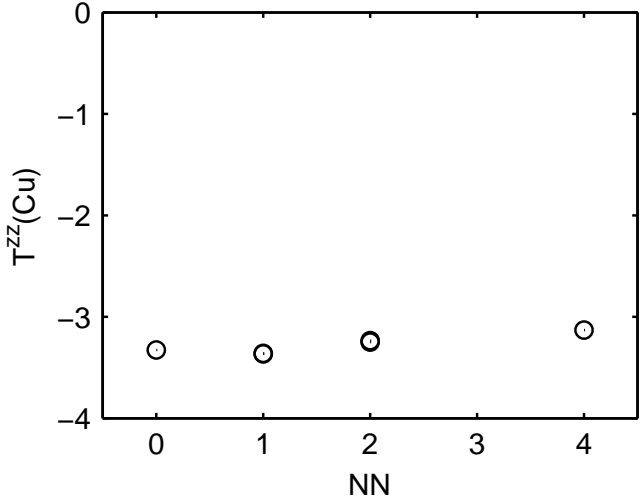}
}
\caption{The dipolar hyperfine field at the planar Cu(2) nucleus $T^{zz}(\mathrm{Cu})$ (in units of~\abmtt) as a function of the number of nearest neighbour copper ions. Note that there is no substantial dependence of $T^{zz}(\mathrm{Cu})$ on NN.}
\label{fig:tzz_cu}
\end{figure}

At variance with the isotropic hyperfine density which for copper exhibits a substantial transferred contribution, $T^{ii}$(Cu) is dominated by the on-site term. The transferred field from NN copper ions is small as can be seen in Fig.~\ref{fig:tzz_cu} where $T^{zz}$(Cu) is plotted as a function of the number of NN.

Again, all obtained values for the different clusters and multiplicities turned out to be proportional to the copper Mulliken spin densities and the same kinds of ansatz that worked for the isotropic case could be used. The results of this analysis can be written as 

\begin{equation}
T^{zz}(\textrm{Cu}_i) = \alpha_{dip}^{\parallel} \rho(\textrm{Cu}_i) + \beta_{dip}^{\parallel} \sum_{j \in NN} \rho(\textrm{Cu}_j)
\label{eq:tzz_cu}
\end{equation}
and

\begin{equation}
T^{ii}(\textrm{O}) = \gamma_{dip}^{ii} \sum_{j \in NN} \rho(\textrm{Cu}_j).
\label{eq:tii_o}
\end{equation}
The fitted parameters $\alpha_{dip}^{\parallel}$, $\beta_{dip}^{\parallel}$, and $\gamma_{dip}^{ii}$ are collected in Table~\ref{tbl:dipolar_param}.

\begin{table*}[htb]
\caption{The values of the fitted parameters of our ansatz~(\ref{eq:tzz_cu}) for the dipolar hyperfine field at Cu(2) and~(\ref{eq:tii_o}) for that at O(2/3) (in units of \abmtt). Note that the dipolar field at the copper nucleus is nearly axially symmetric and $\alpha_{dip}^{\parallel}$ refers to the crystallographic $c$-direction. For the planar oxygens, $\gamma_{dip}^{\parallel}$ ($\gamma_{dip}^{\perp}$) denotes the component in the CuO$_2$ plane along (perpendicular to) the bond axis.}
\begin{center}
\begin{tabular}{ccccc}
\hline
$\alpha_{dip}^{\parallel}$ & $\beta_{dip}^{\parallel}$ & $\gamma_{dip}^{\parallel}$ & $\gamma_{dip}^{\perp}$ & $\gamma_{dip}^{c}$ \\ \hline
$-5.16 \pm 0.01$ & $0.136 \pm 0.004$ & $0.47 \pm 0.01$ & $-0.26 \pm 0.01$ & $-0.21 \pm 0.01$ \\ \hline
\end{tabular}
\label{tbl:dipolar_param}
\end{center}
\end{table*}

%---------------------------------------------------
%---------------------------------------------------
\subsection{Discussion of spin density distribution}
%---------------------------------------------------
%---------------------------------------------------

\label{sec:spin_density}

\begin{figure}[htb]
\resizebox{0.48\textwidth}{!}{%
  \includegraphics{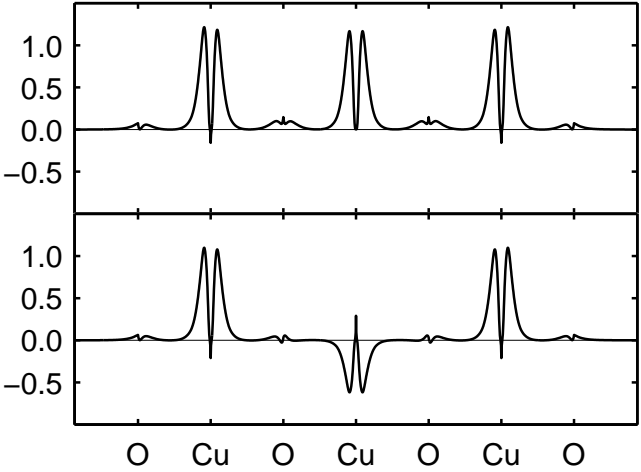}
}
\caption{Spin density distribution (in \abmtt) along the crystallographic axis $a$ in the cluster Cu$_5$O$_{21}$/Cu$_{13}$Y$_{12}$Ba$_{12}$. The upper (lower) panel displays the results from a calculation with spin multiplicity m=6 (m=4).}
\label{fig:spindensity}
\end{figure}

The capability of the cluster approach to control the spin multiplicities enables us to get insight into the details of on-site and transferred spin densities. In Fig.~\ref{fig:spindensity} the difference of spin-up and spin-down densities is shown along a bond axis for multiplicities $m=6$ (upper panel) and $m=4$ (lower panel) calculated with the Cu$_5$O$_{21}$/Cu$_{13}$Y$_{12}$Ba$_{12}$ cluster. The double humps at the coppers reflect the fact that the dominant part of the spin density distribution is related to the projections of the MOs onto the 3d$_{x^2-y^2}$ AOs. The small differences in the heights are due to the finite size of the cluster. The contact density at the two outer coppers is negative (they have only one NN) but positive at the central copper (with four NN). At the oxygens, the spin density resides mainly on the 2p$_{\sigma}$ AO. Again, the tiny peaks at the positions of the nuclei reveal the contact densities which for the two inner oxygen atoms with two NN  is twice as large as for the two outer oxygens with only one NN.
The ``antiferromagnetic'' alignment ($m=4$, lower panel of Fig.~\ref{fig:spindensity}) also exhibits finite size effects. Since there are four copper ions with spin up and only one with spin down, there is an excess of positive spin density in the cluster which prohibits an exact cancellation of $\rho^{\uparrow}$ and $\rho^{\downarrow}$ at the bridging oxygens. The smaller spin density distribution on the central copper atom in the antiferromagnetic case may arise from the removal of the restrictions due to the exclusion principle for the central spin compared to the $m=6$ case.
Note, however, that the electronic structure and spin density distributions for all the clusters are in perfect accordance if the analysis is performed with the individual Mulliken spin densities $\rho(\textrm{Cu})$ as has been demonstrated in the previous subsection (see Figs.~\ref{fig:dcu/rho} and~\ref{fig:do}).

An extrapolation from the finite clusters to an extended system therefore just requires the knowledge of $\rho$(Cu). We assume that the value calculated for the central copper atom in the large cluster, $\rho$(Cu)$=0.68$, is a reasonable estimate. The resulting hyperfine coupling constants ($a_{iso}=0.68 \alpha_{iso}$ and, similarly, for the other couplings) will be given and discussed in the next subsection.

\begin{figure}[htb]
\resizebox{0.48\textwidth}{!}{
\includegraphics{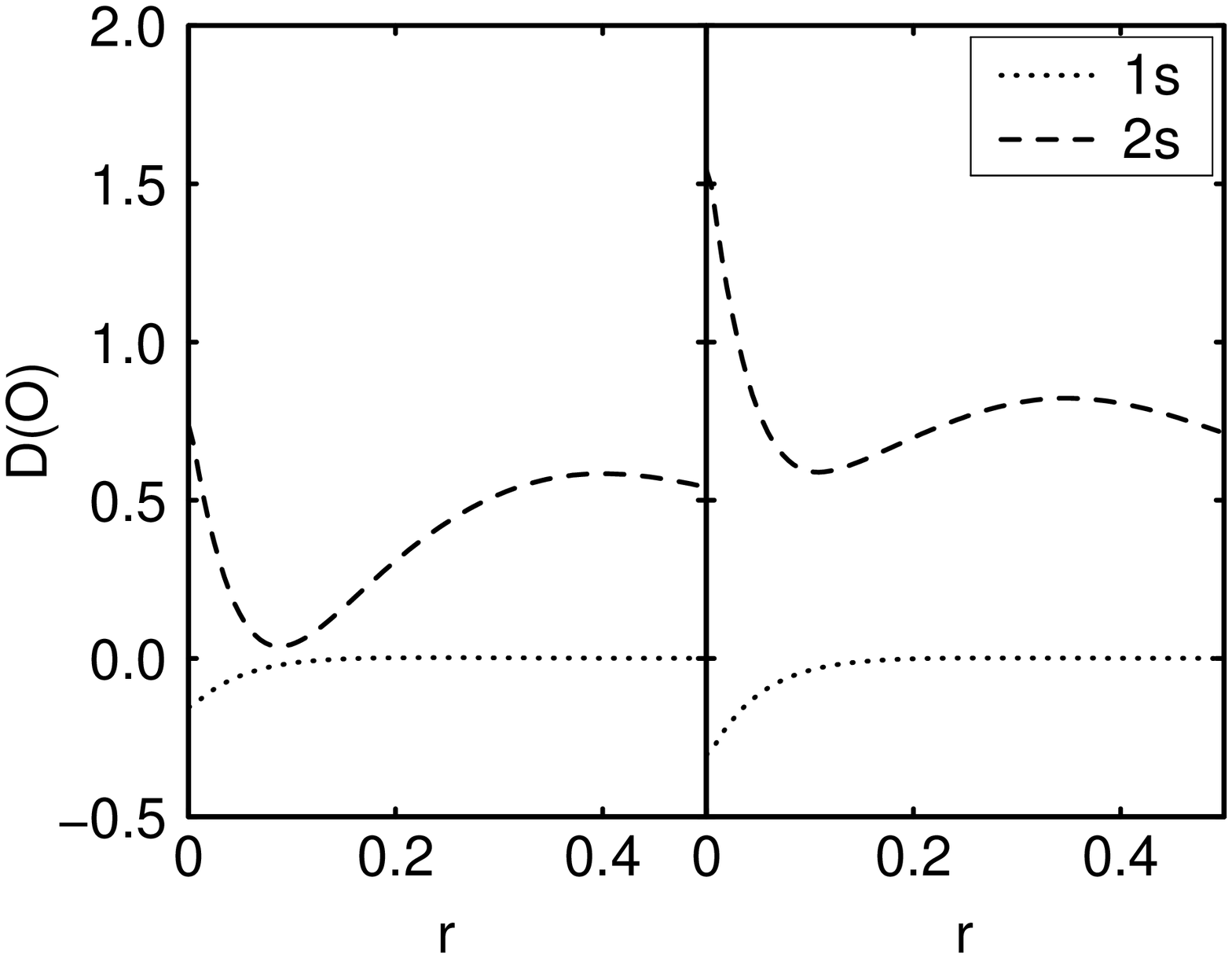}
}
\caption{Radial dependence (along bond direction in units of a$_B$) of $D$ (in \abmtt) subdivided into contributions from MOs with mainly 1s and 2s AO character at a planar oxygen nucleus for the small cluster CuO$_5$/Cu$_5$Y$_{12}$Ba$_{12}$ with $\textrm{NN}=1$ (left panel) and for the large cluster Cu$_5$O$_{21}$/Cu$_{13}$Y$_{12}$Ba$_{12}$ with $\textrm{NN}=2$ (right panel).}
\label{fig:diff_o}
\end{figure}

\begin{figure}[htb]
\resizebox{0.48\textwidth}{!}{
\includegraphics{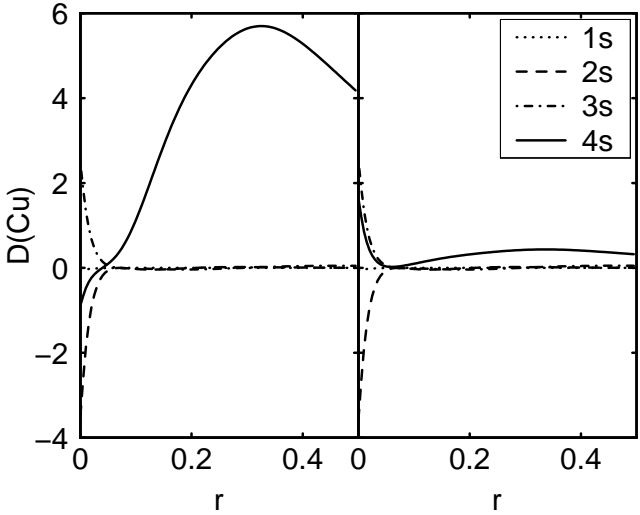}
}
\caption{Radial dependence (along bond direction in units of a$_B$) of $D$ (in a$_B^{-3}$) subdivided into contributions from MOs with mainly 1s, 2s, 3s, and 4s AO character at a planar copper nucleus for the small cluster CuO$_5$/Cu$_5$Y$_{12}$Ba$_{12}$ with $\textrm{NN}=0$ (left panel) and for the large cluster Cu$_5$O$_{21}$/Cu$_{13}$Y$_{12}$Ba$_{12}$ with $\textrm{NN}=4$ (right panel).}
\label{fig:diff_cu}
\end{figure}

To illustrate the mechanisms of spin transfer, the radial dependence of the difference between spin-up and spin-down densities at the oxygen and the copper is shown in Figs.~\ref{fig:diff_o} and~\ref{fig:diff_cu} for the clusters CuO$_5$/Cu$_5$Y$_{12}$Ba$_{12}$ and Cu$_5$O$_{21}$/Cu$_{13}$Y$_{12}$Ba$_{12}$. Owing to overlap and covalence effects the Cu$^{2+}$ ion shares its spin density with the ligand planar oxygens whose spin direction is parallel to that of the local copper moment. The transferred spin is mainly on the O 2p$_{\sigma}$ orbital with a polarization of 8.2\% and polarizes the s-orbitals such that the nucleus sees positive spin density. This can be seen in Fig.~\ref{fig:diff_o} (left panel) where the radial dependence of $D$ near the oxygen nucleus is shown separately for the projections onto the 1s and 2s AOs. The contact densities are $D_{2s}(\textrm{O})=0.737$~\abmtt\ and $D_{1s}(\textrm{O})=-0.154$~\abmtt. If the oxygen is between two NN copper ions (right panel) the contact densities are approximately doubled to $D_{2s}(\textrm{O})=1.543$~\abmtt\ and $D_{1s}(\textrm{O})=-0.304$~\abmtt. The densities $\abs{\psi_{ns}^{\uparrow} (\textrm{O})}^2$ and $\abs{\psi_{ns}^{\downarrow} (\textrm{O})}^2$ are almost identical with values $\abs{\psi_{1s} (\textrm{O})}^2\ = 141\ \abmtm$ and $\abs{\psi_{2s} (\textrm{O})} ^2\ = 6.79\ \abmtm$. Defining the polarizations $f_{ns}$ of the s-like orbitals by

\begin{equation}
D_{ns}(\textrm{O})=\frac{8 \pi}{3}\abs{\psi_{ns}(\textrm{O})}^2 f_{ns}
\end{equation}
we get $f_{2s}=1.3\%$ and $f_{1s}=-0.013\%$ with $\textrm{NN}=1$ but $f_{2s}=2.7\%$ and $f_{1s}=-0.026\%$ for $\textrm{NN}=2$.

The isotropic spin density distributions near the Cu nucleus are shown in Fig.~\ref{fig:diff_cu}, for $\textrm{NN}=0$ ($\textrm{NN}=4$) in the left (right) panel. The corresponding numerical values are summarized in Table~\ref{tbl:polarization_values}. The hyperfine fields transferred from the four NN ions change the polarization of the 4s-orbital from negative to positive.

\begin{table*}[htb]
\caption{Values of the s-like AOs at the Cu (in \abmtt), the isotropic hyperfine densities (in \abmtt), and polarizations $f_{ns}$ for the cluster CuO$_5$/Cu$_5$Y$_{12}$Ba$_{12}$ ($\textrm{NN}=0$) and the cluster Cu$_5$O$_{21}$/Cu$_{13}$Y$_{12}$Ba$_{12}$ ($\textrm{NN}=4$).}
\begin{center}
\begin{tabular}{c|ccc|ccc}
\hline
&\multicolumn{3}{c|}{$\textrm{NN}=0$} &
\multicolumn{3}{c}{$\textrm{NN}=4$}\\ \hline
$n$ & $\abs{\psi_{ns}(\textrm{Cu})}^2$ & $D_{ns}(\textrm{Cu})$ & $f_{ns}[\%]$ & $\abs{\psi_{ns}(\textrm{Cu})}^2$ & $D_{ns}(\textrm{Cu})$ & $f_{ns}[\%]$ \\
1 & 7300 & $-0.0586$ & $-9.59 \times 10^{-5}$ & 7300 & $-0.0561$ & $-9.18 \times 10^{-5}$ \\
2 & 725  & $-3.542$  & $-5.83 \times 10^{-2}$ &  725 & $-3.569$  & $-5.88 \times 10^{-2}$ \\
3 & 107  &   2.506   &   0.280                &  107 &   2.530   &   0.282                \\
4 & 2.47 & $-0.884$  & $-4.27$                & 2.36 &   1.261   &   6.38                 \\ \hline
\end{tabular}
\end{center}
\label{tbl:polarization_values}
\end{table*}

%-------------------------------
%-------------------------------
\subsection{Hyperfine Couplings}
%-------------------------------
%-------------------------------

The results for the magnetic hyperfine couplings for the planar oxygen are represented in Table~\ref{tbl:c_tot}. The spin-orbit contributions to the hyperfine couplings are expected to be small in the case of oxygen. We neglect them and assume that $c_{tot}^{ii}=c_{iso}+c_{dip}^{ii}$. The values are very similar to the ones calculated for La$_2$CuO$_4$ (see Ref.~\cite{bib:physrevb}) with the exception that now $c_{tot}^c$ is somewhat larger than $c_{tot}^{\perp}$.

\begin{table*}[htb]
\caption{Compilation of the magnetic hyperfine coupling constants in units of \abmtt\ for the planar oxygen extrapolated to the infinite crystal.}
\begin{center}
\begin{tabular}{lcccc}
\hline
 & $c_{iso}$ & $c_{tot}^{\parallel}$ & $c_{tot}^{\perp}$ & $c_{tot}^c$ \\ \hline
this work   & 0.646 & 0.966 & 0.469 & 0.503 \\
Experiment$^a$ &        & 1.18 $\pm$ 0.08  & 0.69 $\pm$ 0.08  & 0.81 $\pm$ 0.08  \\ \hline
\multicolumn{5}{l}{$^a$Ref.~\cite{bib:yoshinari}.} \\ \hline
\end{tabular}
\label{tbl:c_tot}
\end{center}
\end{table*}

Experimentally, these couplings cannot be measured directly but they are derived from a combination of various data. The spin shift measured with the external field in direction $i$ is proportional to $c_{tot}^{ii}$ and the static uniform spin susceptibility. The spin-lattice relaxation rates are proportional to the sum of the squares of $c_{tot}^{jj}$ in the directions orthogonal to $i$. Whereas the overall agreement between the theoretical and experimental values is satisfying, it should be noted that the calculated anisotropy, $c_{tot}^{\parallel}/c_{tot}^{\perp} \simeq 2$, is larger than the experimental value of about 1.7.

When comparing the theoretical values with those derived from experiments (see Table~\ref{tbl:c_tot}) it is seen that the latter values are generally higher. The experimental analysis relies on the approximation that the total spin susceptibility $\chi_s$ is isotropic and on the uncertain assumption that it is equally distributed over Cu(2) and Cu(1) sites. (In Ref.~\cite{bib:yoshinari} the value $\chi_s=27.5 \times 10^{-5}$\ emu/mole divided by 2.96 appropriate for the YBa$_2$Cu$_3$O$_{6.96}$ sample was used.) Johnston and Cho~\cite{bib:johnston} reported anisotropic spin susceptibilities $\chi_s^{\parallel}$ ($\chi_s^{\perp}$) of 37.4 (33.5) $\times 10^{-5}$\ emu/mole. Using these values and the spin shifts from Ref.~\cite{bib:yoshinari} the hyperfine coupling constants would be $c_{tot}^{\parallel} = 0.98\ \abmtm$, $c_{tot}^{\perp}=0.58\ \abmtm$, and $c_{tot}^c= 0.61\ \abmtm$.

Yoshinari {\it et al.}~\cite{bib:yoshinari} also found that the hyperfine fields did not appreciably change between samples with an oxygen content of $x=6.6$, 6.8, and 6.96. This is in sharp contrast with recent measurements~\cite{bib:recentwalstedt} in the La$_2$CuO$_4$ system where large differences between the hyperfine fields in undoped and doped (0.15\% Sr) compounds have been observed. 

The calculated values for the isotropic and dipolar hyperfine couplings for the planar copper are given in the left part of Table~\ref{tbl:calc_hyp}. We cannot yet calculate from first principles the spin-orbit contribution which for copper is not negligible. The total values for the on-site densities, $a_{tot}^{ii} = a_{iso} + a_{dip} ^{ii} + a_{so}^{ii}$, however, depend crucially on $a_{so}^{ii}$, partly accidentally as will be shown below. We therefore choose to adopt the estimates given in Ref.~\cite{bib:monienpinesslichter} which are based on the picture of a single missing electron in the copper 3d-shell developed by Bleaney {\it et al.}~\cite{bib:bleaney}. From perturbation theory the hyperfine spin-orbit coupling densities are given by a phenomenological spin-orbit coupling parameter $\lambda$, the energy differences $\Delta E$ between the various 3d-orbitals and $\left<r^{-3}\right>$. Using the ``standard'' values for these quantities which are obtained from a variety of combined experimental and theoretical studies, one obtains $a_{so}^{\parallel}=2.409\ \abmtm$ and $a_{so}^{\perp}=0.427\ \abmtm$. Following the arguments given by Monien, Pines, and Slichter~\cite{bib:monienpinesslichter} who estimate the uncertainties in $\lambda$ and $\Delta E$ to about 20\% and adding 10\% for the possible change in $\left<r^{-3}\right>$, we include the 30\% bounds in the ``errors'' given for $a_{tot}^{\parallel}$ and $a_{tot}^{\perp}$ in Table~\ref{tbl:calc_hyp}.

\begin{table*}[htb]
\caption{The hyperfine couplings for the planar copper atom and comparison with literature values (in units of \abmtt). The errors assigned to the total on-site hyperfine couplings originate from the uncertainties of the spin-orbit couplings as discussed in the text. (Since the dipolar hyperfine interaction tensor is traceless and nearly axially symmetric, $a_{dip}^{\perp} \simeq -\frac{1}{2} a_{dip}^{\parallel}$\ and\ $b_{dip}^{\perp} \simeq -\frac{1}{2} b_{dip}^{\parallel}$).}
\begin{center}
\begin{tabular}{lcccc|cccc}
\hline
& $a_{iso}$ & $b_{iso}$ & $a_{dip}^{\parallel}$ & $b_{dip}^{\parallel}$ &
$a_{so}^{\parallel}$ & $a_{so}^{\perp}$ & $a_{tot}^{\parallel}$ &
$a_{tot}^{\perp}$ \\ \hline
this work & $-1.82$ & 0.49 & $-3.51$ & 0.09 & 2.41 & 0.43 & $-2.93 \pm 0.72$ & $0.36 \pm 0.18$ \\
other$^a$  & $-2.05$ & 0.74 & $-3.67$ & &  2.09 & 0.29 & $-3.64$ & $-0.08$ \\
$^b$       &         & 0.58 &         &        &      && $-3.12$ &   0.40  \\ 
$^c$       &         & 0.71 &         &        &      && $-3.01$ & $-0.18$ \\ 
$^d$       &         & 0.69 &         &        &      && $-2.75$ &   0.50  \\ 
$^e$       &         & 0.62 &         &        &      && $-2.52$ &   0.59  \\ \hline
\multicolumn{9}{l}{$^a$Ref.~\cite{bib:milarice}, $^b$Ref.~\cite{bib:monienpinesslichter}, $^c$Ref.~\cite{bib:walstedtwarren}, $^d$Ref.~\cite{bib:zha}, $^e$Ref.~\cite{bib:imai}.} \\ \hline
\end{tabular}
\label{tbl:calc_hyp}
\end{center}
\end{table*}

In comparing our calculated values with previous estimates and values derived from experiment, we first note that we also get a transferred anisotropic hyperfine coupling $b_{dip}^{\parallel}$ which is 18\% of the value of $b_{iso}$. The value for $b_{iso}$ of $0.49$~\abmtt\ is considerably smaller than that of $0.71$~\abmtt\ obtained in the La$_2$CuO$_4$ compound~\cite{bib:physrevb}. This is due to the buckling of the planar oxygens in the CuO$_2$ plane~\cite{bib:samodiss}. Experiments give constraints on various combinations of the coupling parameters. Measurements of the $^{63}$Cu NMR frequencies in the insulating antiferromagnetic compounds YBa$_2$Cu$_3$O$_6$ and La$_2$CuO$_4$ give similar frequencies of about 90~MHz. This then determines $\abs{a_{tot}^{\perp} - 4 ( b_{iso} + b_{dip}^{\perp} )} = 1.92$~\abmtt\ if an effective localized magnetic moment of $0.66~\mu_B$ is assumed. Our values from Table~\ref{tbl:calc_hyp} yield $1.42$~\abmtt. A second constraint comes from the anisotropy ratio $R$ of the nuclear spin-lattice relaxation rates for fields perpendicular and parallel to the planes which is found to be about 3.7. Assuming antiferromagnetically correlated spins $R$ is given by

\begin{equation}
R = \frac{1}{2} \left[ 1+ \left( \frac{a_{tot}^{\parallel}-4 \left( b_{iso}+b_{dip}^{\parallel}\right)}{a_{tot}^{\perp}-4 \left( b_{iso}+b_{dip}^{\perp} \right)} \right)^2 \right].
\label{eq:ratio}
\end{equation}

Inserting our calculated hyperfine couplings into Eq.~(\ref{eq:ratio}) yields values for $R$ which are far too large. An inspection of Eq.~(\ref{eq:ratio}), however, shows that $R$ depends sensitively on $a_{tot}^{\perp}$. In the nominator both terms add since $a_{tot}^{\parallel} < 0$. In the denominator, however, a cancellation takes place since $a_{tot}^{\perp} > 0$, and the relative values of the terms are important. Due to the accidental near cancellation of $a_{iso}$ and $a_{dip}^{\perp}$ ($a_{iso}+a_{dip}^{\perp}=-0.068$~\abmtt), the value of $a_{so}^{\perp}$ is crucial for $a_{tot}^{\perp}$, as has been shown by Stoll {\it et al.}~\cite{bib:klosters} for La$_2$CuO$_4$ where, in addition, the spin-orbit contributions to the hyperfine field at the copper nucleus are strongly influenced by Sr$^{2+}$ dopants. Therefore the calculated couplings are inadequate in predicting the anisotropy of the Cu relaxation rates until more reliable values for $a_{so}^{\perp}$ are available.

A third experimental constraint comes from the fact that the copper magnetic shift measured for fields perpendicular to the plane is independent of temperature and therefore assigned entirely to contributions from the chemical shift leaving a vanishing spin shift $^{63}K_s^{\parallel}$. This can happen if $a_{tot}^{\parallel}+4(b_{iso}+b_{dip}^{\parallel})=0$. Our calculated values in Table~\ref{tbl:calc_hyp} give $-0.61$~\abmtt\ and a value of $a_{so}^{\parallel} \simeq 3$~\abmtt\ would be required for a perfect cancellation. It should be emphasized, however, that $^{63}K_s^{\parallel}$ is found to be strictly temperature independent for most of the cuprates irrespective of the doping level. This is a remarkable experimental fact and an accidental cancellation of the various terms for several substances and doping levels is completely unexpected from our theoretical calculations of hyperfine parameters.

%--------------------------------
%--------------------------------
\section{Summary and Conclusions}
%--------------------------------
%--------------------------------

\label{sec:conclusion}

The local electronic structure of YBa$_2$Cu$_3$O$_7$ has been investigated by first-principles methods. Using clusters of various sizes, which attempt to incorporate the important features of the CuO$_2$-plane, the EFGs at the planar copper and oxygen sites were calculated with all-electron calculations and sufficiently large basis sets. The calculated values are in good agreement with the experiments. An inspection of Table~\ref{tbl:detailedefg_cupm4}, where we examined in detail the various contributions to the EFG at the copper nucleus, demonstrates that the traditional interpretation of EFG values using Sternheimer shielding and antishielding factors is highly questionable.

The magnetic hyperfine properties in the planes have been determined very accurately by comparing the results from calculations with different spin multiplicities. It was shown that the transferred hyperfine fields at both Cu(2) and O(2/3) are effectively due only to the nearest neighbour copper ions, the contributions from further distant neighbours being marginal. This finding has important consequences since it shows that the assumption of Zha, Barzykin, and Pines~\cite{bib:zha} has no microscopic justification. The variant temperature dependencies of the Cu(2) and O(2/3) nuclear spin lattice relaxation rates can then only be explained by antiferromagnetic fluctuations which are commensurate. It can be argued that the neutron scattering which exhibits incommensurate fluctuations is operating on a much faster time scale than the NMR experiments or is probing periodic arrays of domain walls.

The calculated values for the hyperfine coupling constants are in reasonable agreement with the values expected from experimental data. As the discussion in Sec.~\ref{sec:mag_hyp} demonstrated, the interpretation of the data is made difficult by the uncertainty of the contributions of the spin-orbit interaction. In this respect, improved quantum chemical calculations of the spin-orbit coupling would be desirable.

\vspace{0.5cm}

The initial studies for this work were done in the diploma work of W.O. Dijkstra whose careful analysis is gratefully acknowledged. We express our gratitude to J. Haase, M. Mali, J. Roos, H.U. Suter, and R.E. Walstedt for enlightening discussions. This work is partially supported by the Swiss National Science Foundation.


\begin{thebibliography}{99}

\bibitem{bib:bednorzmuller}
        J.G. Bednorz, K.A. M\"uller, Z. Phys. {\bf 64}, 189 (1986).

\bibitem{bib:reviews}
For reviews, see: 
        D. Brinkmann, M. Mali, {\it NMR Basic Principles and Progress}
        (Springer, Heidelberg, 1994, Vol. 31, p. 171);
        C.P. Slichter,
        in {\it Strongly Correlated Electronic Materials}, edited by K.S. Bedell {\it et al.},
        (Addison-Wesley 1994);
        A. Rigamonti, F. Borsa, P. Caretta, Rep. Prog. Phys. {\bf 61}, 1367 (1998).

\bibitem{bib:milarice} F. Mila, T.M. Rice,
        Physica C {\bf 157}, 561 (1989).

\bibitem{bib:shastry} B.S. Shastry,
        \prl\ {\bf 63}, 1288 (1989).

\bibitem{bib:zha} Y. Zha, V. Barzykin, D. Pines, 
        \prb\ {\bf 54}, 7561 (1996).

\bibitem{bib:walstedtshastry}
        R.E. Walstedt, B.S. Shastry, S.-W. Cheong,
        Phys. Rev. Lett. {\bf 72}, 3610 (1994).

\bibitem{bib:martindale}
        J.A. Martindale, P.C. Hammel, W.L. Hults, J.L. Smith,
        Phys. Rev. B {\bf 57}, 11769 (1998).

\bibitem{bib:penningtonyugorny}
        C.H. Pennington, S. Yu, K.R. Gorny, M.J. Buoni, W.L. Hults, J.L. Smith,
        \prb\ {\bf 63}, 054513 (2001).

\bibitem{bib:das} T.P. Das in {\it Electronic Properties of Solids Using Cluster
  Methods}, edited by T.A. Kaplan, S.D. Mahanti (Plenum Press, New York, 1995).

\bibitem{bib:das1}
        N. Sahoo, S. Markert, T.P. Das, K. Nagamine, 
        \prb\ {\bf 41}, 220 (1990).

\bibitem{bib:das2}
        S.B. Sulaiman, N. Sahoo, T.P. Das, O. Donzelli,
        \prb\ {\bf 45}, 7383 (1992).

\bibitem{bib:schwarz1}
        K. Schwarz, C. Ambrosch-Draxl, P. Blaha, \prb\ {\bf 42}, 2051 (1990).

\bibitem{bib:schwarz2}
        C. Ambrosch-Draxl, P. Blaha, K. Schwarz, \prb\ {\bf 44},
        5141 (1991).

\bibitem{bib:yu}
        J. Yu, A.J. Freeman, R. Podloucky, P. Herzig, P. Weinberger,
        \prb\ {\bf 43}, 532 (1991). 

\bibitem{bib:winter}
        N.W. Winter, C.I. Merzbacher, C.E. Violet, Appl. Spec. Rev.
        {\bf 28}, 123 (1993).

\bibitem{bib:physicac}
        P. H\"usser, E. Stoll, H.U. Suter, P.F. Meier,
        Physica C {\bf 294} 217 (1998).

\bibitem{bib:physrevb}
        P. H\"usser, H.U. Suter, E.P. Stoll, P.F. Meier,
        \prb\ {\bf 61}, 1567 (2000).

\bibitem{bib:unitcell}
        P. Bordet, C. Chaillout, J.J. Capponi, J. Chenavas, 
        M. Marezio, Nature {\bf 327} 687 (1987).

\bibitem{bib:g98}
 M.J. Frisch, {\it et al.}, Gaussian 98, Revision A.5, (Gaussian, Inc., Pittsburgh PA, 1998).

\bibitem{bib:pseudopotentials}
Ba$^{2+}$: W.R. Wadt, P.J. Hay, J. Chem. Phys. {\bf 82}, 284 (1985);
Y$^{3+}$: P.J. Hay,  W.R. Wadt, J. Chem. Phys. {\bf 82}, 270 (1985);
Cu$^+$: P. Fuentealba, H. Stoll, L. v. Szentp\'aly, P. Schwerdtfeger, H. Preuss,
        J. Phys. B: At. Mol. Phys. {\bf 16}, L323 (1983);
Cu$^{2+}$: R.L. Martin (private communication).

\bibitem{bib:becke1}
        A.D. Becke, \pra\ {\bf{38}}, 3098 (1988).

\bibitem{bib:becke2}  A.D. Becke, J. Chem. Phys. {\bf 88}, 2547 (1988).

\bibitem{bib:LYP}  C. Lee, W. Yang, R.G. Parr,
        \prb\ {\bf 37}, 785 (1988).

\bibitem{bib:stoll97}
        E.P. Stoll (unpublished).

\bibitem{bib:bleaney}{
B. Bleaney, K.D. Bowers, M.H.L. Pryce, Proc. R. Soc. London, Ser. A {\bf 228}, 166 (1955).}

\bibitem{bib:pennington}
        C.H. Pennington, D.J. Durand, C.P. Slichter, J.P. Rice, E.D.
        Bukowski, D.M. Ginsberg, \prb\ {\bf 39}, 2902 (1989).

\bibitem{bib:sternheimer}{R.M. Sternheimer, Z. Naturforsch. {\bf 41a}, 35 (1985).}

\bibitem{bib:epstoll_efg}
        E.P. Stoll (to be published).

\bibitem{bib:takigawa}
        M. Takigawa, P.C. Hammel, R.H. Heffner, Z. Fisk, K.C. Ott, J.D. Thompson, \prl\ {\bf 63}
        1865 (1989).

\bibitem{bib:yoshinari}
        Y. Yoshinari, H. Yasuoka, Y. Ueda, K. Koga, K. Kosuge,
        J. Phys. Soc. Japan {\bf 59}, 3698 (1990).

\bibitem{bib:johnston}
        D.C. Johnston, J.H. Cho, \prb\ {\bf 42}, 8710 (1990).

\bibitem{bib:recentwalstedt}
        R.E. Walstedt, S.-W. Cheong, \prb\ {\bf 64}, 014404 (2001).

\bibitem{bib:monienpinesslichter}
        H. Monien, D. Pines, C.P. Slichter,
        \prb\ {\bf 41}, 11120 (1990).

\bibitem{bib:walstedtwarren}
        R.E. Walstedt, W.W. Warren,
        Science {\bf 248}, 1082 (1990).

\bibitem{bib:imai}
        T. Imai, J. Phys. Soc. Japan {\bf 59}, 2508 (1990).

\bibitem{bib:samodiss}
        S. Pliber\v{s}ek (to be published).

\bibitem{bib:klosters}
        E.P. Stoll, S. Pliber\v{s}ek, S. Renold, T.A. Claxton, and P.F. Meier,
        J. Supercond. {\bf 13}, 971 (2000).

\end{thebibliography}
\end{document}